\documentclass[paper]{JHEP3}
\def\mhat{\widehat{m}}
\def\beq{\begin{equation}}
\def\eeq{\end{equation}}
\def\beqn{\begin{eqnarray}}
\def\eeqn{\end{eqnarray}}
\preprint{Cavendish--HEP--06/15}
\title{Addendum to ``Distinguishing Spins in Decay Chains at the Large Hadron Collider''%
\footnote{Work supported in part by the UK Particle Physics and
Astronomy Research Council.}}
\author{Christiana Athanasiou$^1$, Christopher G.\ Lester$^2$,
Jennifer M.\ Smillie$^3$ and Bryan R.\ Webber$^4$\\
  Cavendish Laboratory, University of Cambridge,\\
  JJ Thomson Avenue, Cambridge CB3 0HE, U.K.\\
  $^1$E-mail: \email{ca274@cam.ac.uk}\\
  $^2$E-mail: \email{lester@hep.phy.cam.ac.uk}\\
  $^3$E-mail: \email{smillie@hep.phy.cam.ac.uk}\\
  $^4$E-mail: \email{webber@hep.phy.cam.ac.uk}}
\abstract{We extend our earlier study of spin correlations in
the decay chain $D\to Cq$, $C\to Bl^{\rm near}$, $B\to Al^{\rm far}$,
where $A,B,C,D$ are new particles with known masses but
undetermined spins, $l^{\rm near}$ and $l^{\rm far}$ are
opposite-sign same-flavour charged leptons and $A$ is invisible.
Instead of looking at the observable 2- and 3-particle invariant
mass distributions separately, we compare the full three-dimensional
phase space distributions for all possible spin assignments of the
new particles, and show that this enhances their distinguishability
using a quantitative measure known as the Kullback-Leibler distance.}

\begin{document}
\section{Introduction}
In the recent paper \cite{Athanasiou:2006ef}, to which we refer the reader for
motivation, notation and relevant references, we examined the
distinguishability of different spin assignments in the decay chain
 $D\to Cq$, $C\to Bl^{\rm near}$, $B\to Al^{\rm far}$,
where $A,B,C,D$ are new particles with known masses but
undetermined spins, $l^{\rm near}$ and $l^{\rm far}$ are
opposite-sign same-flavour charged leptons and $A$ is invisible.
This was done by comparing separately the invariant mass distributions
of the three observable two-body combinations: dileptons ($m_{ll}$), quark- or
antiquark-jet plus positive lepton ($m_{jl^+}$), and jet plus
negative lepton ($m_{jl^-}$).\footnote{The three-body invariant mass $m_{jll}$
was also studied but this is not independent of the two-body masses.}

If $P(m|S)$ represents the normalized probability distribution
of any one of these three invariant masses predicted by spin assignment
$S$, and $T$ is the true spin configuration, then a measure of the
improbability of $S$ is provided by the {\em Kullback-Leibler distance}
\beq\label{eq:KLdist}
{\rm KL}(T,S) = \int_m\log\left(\frac{P(m|T)}{P(m|S)}\right)P(m|T) d m\;.
\eeq
In particular, the number $N$ of events required to disfavour hypothesis
$S$ by a factor of $1/R$ under ideal conditions, assuming equal prior
probabilities of $S$ and $T$, would be
\beq\label{eq:NST}
N \sim \frac{\log R}{{\rm KL}(T,S)}\;.
\eeq
By ideal conditions we mean isolation of the decay chain with no
background and perfect resolution. Therefore $N$ sets a lower limit on the
number of events that would be needed in real life. The results for $R=1000$
are shown in tables~\ref{tab:AllSUSY.dat}-\ref{tab:MjlmSUSY.dat},
reproduced for convenience from \cite{Athanasiou:2006ef}, where a
discussion of them can be found.  Recall that the notation used is
$DCBA$ with F for fermion, S for scalar, V for vector, so that squark
decay in SUSY is SFSF and excited quark decay in UED is FVFV. Mass
spectra I and II are SUSY- and UED-like respectively (see 
\cite{Athanasiou:2006ef} for details).

\section{Three-dimensional analysis}
To extract the most information from the data we should compare the predictions
of different spin assigments with the full probability distribution in the
three-dimensional space of $m_{ll}$,  $m_{jl^+}$ and $m_{jl^-}$. The
ambiguity between near and far leptons means that this given by
\beqn\label{P3d}
P(m_{ll},m_{jl^+},m_{jl^-}) &=& \frac 12 f_q \left[P_2(m_{ll},m_{jl^+},m_{jl^-})+P_1(m_{ll},m_{jl^-},m_{jl^+})\right]
\nonumber\\
&+&  \frac 12 f_{\bar q} \left[P_1(m_{ll},m_{jl^+},m_{jl^-})+P_2(m_{ll},m_{jl^-},m_{jl^+})\right]\;,
\eeqn
where $f_q$ and $f_{\bar q}=1-f_q$ are the fractions of quark- and
antiquark-like objects $D$ initiating the decay chain and we use
$P_{1,2}(m_{ll},m_{jl}^{\rm near},m_{jl}^{\rm far})$ on the right-hand side,
assuming both leptons are left-handed, otherwise $f_q$ and $f_{\bar q}$
are interchanged. The subscripts 1 and 2 refer to processes 1 and 2 defined in
\cite{Athanasiou:2006ef} and the factors of one-half enter because
$P_{1,2}$ are both normalized to unity. 

Instead of trying to evaluate the three-dimensional generalization of the integral
in eq.~(\ref{eq:KLdist}) analytically, it is convenient to perform a Monte Carlo
integration. If we generate $m_{ll}$, $m_{jl}^{\rm near}$ and $m_{jl}^{\rm far}$
according to phase space, the weight to be assigned to the configuration
$l^{\rm near}=l^+$, $l^{\rm far}=l^-$ is
\beq
P_{+-}(m_{ll},m_{jl}^{\rm near},m_{jl}^{\rm far})
= \frac 12 \left[f_q P_2(m_{ll},m_{jl}^{\rm near},m_{jl}^{\rm far})
           + f_{\bar q} P_1(m_{ll},m_{jl}^{\rm near},m_{jl}^{\rm far})\right]
 \eeq
 while that for $l^{\rm near}=l^-$, $l^{\rm far}=l^+$ is
\beq
P_{-+}(m_{ll},m_{jl}^{\rm near},m_{jl}^{\rm far})
= \frac 12 \left[f_q P_1(m_{ll},m_{jl}^{\rm near},m_{jl}^{\rm far})
           + f_{\bar q} P_2(m_{ll},m_{jl}^{\rm near},m_{jl}^{\rm far})\right]\;.
 \eeq
In the former case, since the distinction between  $l^{\rm near}$ and $l^{\rm far}$ is lost in the data
(except when interchanging them gives a point outside phase space),
we must use eq.~(\ref{P3d}) with $l^+=l^{\rm near}$, $l^-=l^{\rm far}$ in the logarithmic factor of the KL-distance, i.e. the contribution is
\beq
\log\left(\frac{
P_{+-}(m_{ll},m_{jl}^{\rm near},m_{jl}^{\rm far}|T) +
P_{-+}(m_{ll},m_{jl}^{\rm far},m_{jl}^{\rm near}|T)}{
P_{+-}(m_{ll},m_{jl}^{\rm near},m_{jl}^{\rm far}|S) +
P_{-+}(m_{ll},m_{jl}^{\rm far},m_{jl}^{\rm near}|S)}\right)
P_{+-}(m_{ll},m_{jl}^{\rm near},m_{jl}^{\rm far}|T)\;.
\eeq
Similarly from the configuration $l^{\rm near}=l^-$, $l^{\rm far}=l^+$ we get the contribution
\beq
\log\left(\frac{
P_{-+}(m_{ll},m_{jl}^{\rm near},m_{jl}^{\rm far}|T) +
P_{+-}(m_{ll},m_{jl}^{\rm far},m_{jl}^{\rm near}|T)}{
P_{-+}(m_{ll},m_{jl}^{\rm near},m_{jl}^{\rm far}|S) +
P_{+-}(m_{ll},m_{jl}^{\rm far},m_{jl}^{\rm near}|S)}\right)
P_{-+}(m_{ll},m_{jl}^{\rm near},m_{jl}^{\rm far}|T)\;.
\eeq
Denoting the sum of these two contributions at the $i$th phase space point by ${\rm KL}_i(T,S)$,
and summing over $M$ such points, we have as $M\to \infty$
\beq
\frac{M\log R}{\sum_i {\rm KL}_i(T,S)}\to N\;,
\eeq
which is the Monte Carlo equivalent of eq.~(\ref{eq:NST}).
Results for $R=1000$ and $M=5\times 10^7$ are shown in
table~\ref{tab:M3dim.dat}. By comparing with
tables~\ref{tab:AllSUSY.dat}-\ref{tab:MjlmSUSY.dat},
we see that, as might be expected, the three-dimensional analysis
achieves a discrimination that is better than that of a one-dimensional
analysis applied to any single invariant mass distribution.  This could be
particularly useful in difficult cases like that of distinguishing between
SFSF (SUSY) and FVFV (UED). 

\section*{Acknowledgements}
We thank Sabine Kraml and members of the Cambridge Supersymmetry Working Group for helpful comments.

%%%%%%%%%%%%%%%%%%%%%%%%%%%%%%%%%%%%%%%%%%%%%%%%%%%%%%%%%%%%%%%%%%%%%%
%%%%%%%%%%%%%%%%%%%%%%%%% bibliography %%%%%%%%%%%%%%%%%%%%%%%%%%%%%%%
%%%%%%%%%%%%%%%%%%%%%%%%%%%%%%%%%%%%%%%%%%%%%%%%%%%%%%%%%%%%%%%%%%%%%%

\clearpage
\begin{table}
\begin{tabular}[]{@{}r@{$\,$}|@{$\,$}r@{$\;$}r@{$\;$}r@{$\;$}r@{$\;$}r@{$\;$}r@{}}
 \small (a) $\;$ & \small SFSF & \small FVFV & \small FSFS & \small FVFS & \small FSFV & \small SFVF \\
\hline \small
\small SFSF  & \small $\infty$	& \small 60486	& \small 23	& \small 148	& \small
15608	& \small 66 \\ 
\small FVFV  & \small 60622	& \small $\infty$	& 22	& \small 164	& \small
6866	& \small 62 \\ 
\small FSFS  & \small 36	& \small 34	& \small $\infty$	& 16	& \small
39	& \small 266 \\ 
\small FVFS  & \small 156	& \small 173	& \small 11	& \small $\infty$	&
130	& \small 24 \\ 
\small FSFV  & \small 15600	& \small 6864	& \small 25	& \small 122	& \small
$\infty$	& \small 76 \\ 
\small SFVF  & \small 78	& \small 73	& \small 187	& \small 27	& \small
90	& $\infty$ 
\end{tabular}
\hspace{0.2cm}
\begin{tabular}[]{@{}r@{$\,$}|@{$\,$}r@{$\;$}r@{$\;$}r@{$\;$}r@{$\;$}r@{$\;$}r@{}}
\small (b) $\;$ & \small  SFSF & \small  FVFV & \small  FSFS & \small  FVFS & \small  FSFV & \small  SFVF \\
\hline 
\small SFSF  & \small  $\infty$	& \small  3353	& \small  23	& \small  304	& \small
427	& \small  80 \\ \small  
FVFV  & \small  3361	& \small  $\infty$	& \small  27	& \small  179	& \small
232	& \small  113 \\ \small  
FSFS  & \small  36	& \small  44	& \small  $\infty$	& \small  20	& \small
22	& \small  208 \\ \small  
FVFS  & \small  313	& \small  184	& \small  14	& \small  $\infty$	& \small
13077	& \small  35 \\ \small  
FSFV  & \small  436	& \small  236	& \small  15	& \small  12957	& \small  $\infty$
& \small  39 \\ \small  
SFVF  & \small  89	& \small  126	& \small  134	& \small  38	& \small  42	&
\small  $\infty$ 
\end{tabular}
\caption{The number of events needed to disfavour the column model with respect to the row 
 model by a factor of 0.001, assuming the data to come from the row model, for the
 $\mhat_{ll}^2$ distribution: (a) mass spectrum I and (b) mass spectrum II.}
\label{tab:AllSUSY.dat}
\end{table}

\begin{table}
\begin{tabular}[]{@{}r@{$\,$}|@{$\,$}r@{$\;$}r@{$\;$}r@{$\;$}r@{$\;$}r@{$\;$}r@{}}
 \small (a) $\;$ & \small  SFSF & \small  FVFV & \small  FSFS & \small  FVFS & \small  FSFV & \small  SFVF \\
\hline 
\small SFSF  & \small  $\infty$	& \small  1059	& \small  205	& \small  1524	& \small  758	& \small  727 \\ \small 
FVFV  & \small  1090	& \small  $\infty$	& \small  404	& \small  3256	& \small  4363	& \small  1746 \\ \small 
FSFS  & \small  278	& \small  554	& \small  $\infty$	& \small  418	& \small  741	& \small  870 \\ \small 
FVFS  & \small  1605	& \small  3242	& \small  345	& \small  $\infty$	& \small  1256	& \small  2365 \\ \small 
FSFV  & \small  749	& \small  4207	& \small  507	& \small  1212	& \small  $\infty$	& \small  1803 \\ \small 
SFVF  & \small  813	& \small  1821	& \small  751	& \small  2415	& \small  1888	& \small  $\infty$
\end{tabular}
\hspace{0.2cm}
\begin{tabular}[]{@{}r@{$\,$}|@{$\,$}r@{$\;$}r@{$\;$}r@{$\;$}r@{$\;$}r@{$\;$}r@{}}
\small (b) $\;$ & \small  SFSF & \small  FVFV & \small  FSFS & \small  FVFS & \small  FSFV
& \small  SFVF \\ \hline \small  
SFSF  & \small  $\infty$	& \small  3006	& \small  958	& \small  6874	& \small  761	& \small  1280 \\ \small 
FVFV  & \small  2961	& \small  $\infty$	& \small  4427	& \small  1685	& \small  2749	& \small  3761 \\ \small 
FSFS  & \small  914	& \small  4201	& \small  $\infty$	& \small  743	& \small  9874	& \small  4877 \\ \small 
FVFS  & \small  6716	& \small  1699	& \small  752	& \small  $\infty$	& \small  656	& \small  1306 \\ \small 
FSFV  & \small  720	& \small  2666	& \small  10279	& \small  649	& \small  $\infty$	& \small  4138 \\ \small 
SFVF  & \small  1141	& \small  3517	& \small  5269	& \small  1276	& \small  4259	& \small  $\infty$
\end{tabular}
\caption{As in table \ref{tab:AllSUSY.dat}, for the $\mhat_{jl+}^2$ distribution.}
%  (a) mass spectrum I (figure 7a) and (b) mass spectrum II (figure 7b).}
\label{tab:KjlpSUSY.dat}
\end{table}

\begin{table}
\begin{tabular}[]{@{}r@{$\,$}|@{$\,$}r@{$\;$}r@{$\;$}r@{$\;$}r@{$\;$}r@{$\;$}r@{}}
 \small (a) $\;$ & \small  SFSF & \small  FVFV & \small  FSFS & \small  FVFS & \small  FSFV & \small  SFVF \\ 
\hline \small 
SFSF  & \small  $\infty$	& \small  1058	& \small  505	& \small  769	& \small  816	& \small  619 \\ \small 
FVFV  & \small  1090	& \small  $\infty$	& \small  541	& \small  5878	& \small  4821	& \small  445 \\ \small 
FSFS  & \small  565	& \small  714	& \small  $\infty$	& \small  1032	& \small  741	& \small  2183 \\ \small 
FVFS  & \small  799	& \small  6435	& \small  882	& \small  $\infty$	& \small  2742	& \small  510 \\ \small 
FSFV  & \small  806	& \small  4641	& \small  507	& \small  2451	& \small  $\infty$	& \small  413 \\ \small 
SFVF  & \small  692	& \small  541	& \small  2272	& \small  576	& \small  521	& \small  $\infty$
\end{tabular}
\hspace{0.2cm}
\begin{tabular}[]{@{}r@{$\,$}|@{$\,$}r@{$\;$}r@{$\;$}r@{$\;$}r@{$\;$}r@{$\;$}r@{}}
 \small (b) $\;$ & \small  SFSF & \small  FVFV & \small  FSFS & \small  FVFS & \small  FSFV & \small  SFVF \\ 
\hline \small 
SFSF  & \small  $\infty$	& \small  3037	& \small  689	& \small  8633	& \small  925	& \small  967 \\ \small 
FVFV  & \small  2985	& \small  $\infty$	& \small  2271	& \small  1431	& \small  4368	& \small  2527 \\ \small 
FSFS  & \small  707	& \small  2297	& \small  $\infty$	& \small  526	& \small  9874	& \small  5004 \\ \small 
FVFS  & \small  8392	& \small  1450	& \small  525	& \small  $\infty$	& \small  653	& \small  843 \\ \small 
FSFV  & \small  924	& \small  4287	& \small  10279	& \small  640	& \small  $\infty$	& \small  4036 \\ \small 
SFVF  & \small  1047	& \small  2693	& \small  5213	& \small  870	& \small  4041	& \small  $\infty$
\end{tabular}
\caption{As in table \ref{tab:AllSUSY.dat}, for the $\mhat_{jl-}^2$ distribution.}
%  (a) mass spectrum I (figure 8a) and (b) mass spectrum II (figure 8b).}
\label{tab:MjlmSUSY.dat}
\end{table}

\begin{table}
\begin{tabular}[]{@{}r@{$\,$}|@{$\,$}r@{$\;$}r@{$\;$}r@{$\;$}r@{$\;$}r@{$\;$}r@{}}
 \small (a) $\;$ & \small  SFSF & \small  FVFV & \small  FSFS & \small  FVFS & \small  FSFV & \small  SFVF \\ 
\hline \small 
SFSF  & \small  $\infty$	& \small  455	& \small  21	& \small  47	& \small  348	& \small  55 \\ \small 
FVFV  & \small  474	& \small  $\infty$	& \small  21	& \small  54	& \small  1387	& \small  55 \\ \small 
FSFS  & \small  33	& \small  34	& \small  $\infty$	& \small  13	& \small  39	& \small  188 \\ \small 
FVFS  & \small  55	& \small  67	& \small  10	& \small  $\infty$	& \small  54	& \small  19 \\ \small 
FSFV  & \small  341	& \small  1339	& \small  25	& \small  45	& \small  $\infty$	& \small  66 \\ \small 
SFVF  & \small  62	& \small  64	& \small  143	& \small  19	& \small  79	& \small  $\infty$
\end{tabular}
\hspace{0.2cm}
\begin{tabular}[]{@{}r@{$\,$}|@{$\,$}r@{$\;$}r@{$\;$}r@{$\;$}r@{$\;$}r@{$\;$}r@{}}
 \small (b) $\;$ & \small  SFSF & \small  FVFV & \small  FSFS & \small  FVFS & \small  FSFV & \small  SFVF \\ 
\hline \small 
SFSF  & \small  $\infty$	& \small  1053	& \small  21	& \small  230	& \small  194	& \small  63 \\ \small 
FVFV  & \small  1047	& \small  $\infty$	& \small  27	& \small  135	& \small  190	& \small  90 \\ \small 
FSFS  & \small  33	& \small  42	& \small  $\infty$	& \small  19	& \small  22	& \small  175 \\ \small 
FVFS  & \small  242	& \small  140	& \small  13	& \small  $\infty$	& \small  332	& \small  33 \\ \small 
FSFV  & \small  189	& \small  194	& \small  14	& \small  315	& \small  $\infty$	& \small  37 \\ \small 
SFVF  & \small  66	& \small  95	& \small  118	& \small  35	& \small  41	& \small  $\infty$
\end{tabular}
\caption{As in table \ref{tab:AllSUSY.dat}, for the combined three-dimensional distribution.}
%  (a) mass spectrum I and (b) mass spectrum II.}
\label{tab:M3dim.dat}
\end{table}
\end{document}